\documentclass[aps,prb,twocolumn,superscriptaddress,showkeys]{revtex4-1}
\usepackage[utf8]{inputenc}
\usepackage{graphicx}
\usepackage{bm}
\usepackage{color}
\usepackage{natbib}
\usepackage{amsmath}
\usepackage{amssymb}
\usepackage[utf8]{inputenc}

\begin{document}
\title{Spin-orbit coupling   in elemental two--dimensional materials}

\author{Marcin Kurpas}
\email{marcin.kurpas@us.edu.pl}
\affiliation{
 Institute of Physics, University of Silesia in Katowice, 41-500 Chorz\'{o}w, Poland}
\author{Paulo E. Faria~Junior}
\affiliation{Institute for Theoretical Physics, University of Regensburg, Regensburg 93040, Germany}
\author{Martin Gmitra}
\affiliation{Institute of Physics, P. J. \v{S}af\'{a}rik University in Ko\v{s}ice, Park Angelinum 9, 040 01 Ko\v{s}ice, Slovakia
}
\author{Jaroslav Fabian}
\affiliation{Institute for Theoretical Physics, University of Regensburg, Regensburg 93040, Germany}
\begin{abstract}
    The fundamental spin-orbit coupling and spin mixing in graphene and rippled honeycomb lattice materials silicene, germanene, stanene, blue phosphorene, arsenene, antimonene, and bismuthene is investigated from first principles. 
    The intrinsic spin-orbit coupling in graphene is revisited using multi-band $k\cdot p$ theory, showing the
    presence of non-zero spin mixing in graphene despite the mirror symmetry. However, the spin mixing itself does not lead to the the Elliott-Yafet spin relaxation mechanism, unless the mirror symmetry is broken by external factors. For other aforementioned elemental materials we present
    the spin-orbit splittings at relevant symmetry points, as well as
    the spin admixture $b^2$ as a function of energy close to the band extrema or Fermi levels. We find that spin-orbit coupling scales as the square of the atomic number Z, as expected for valence electrons in atoms. For isolated bands, it is found that $b^2\sim Z^4$. The spin-mixing parameter also exhibits giant anisotropy which, 
    to a large extent, can be controlled by tuning the Fermi level. Our results for $b^2$ can be directly transferred to spin relaxation time  due to the Elliott-Yafet mechanism, and therefore provide an estimate of the upper limit for spin lifetimes in materials with space inversion center. 
    
\end{abstract}
\maketitle
\section{Introduction}
\begin{figure}
    \centering
    \includegraphics[width=0.95\columnwidth]{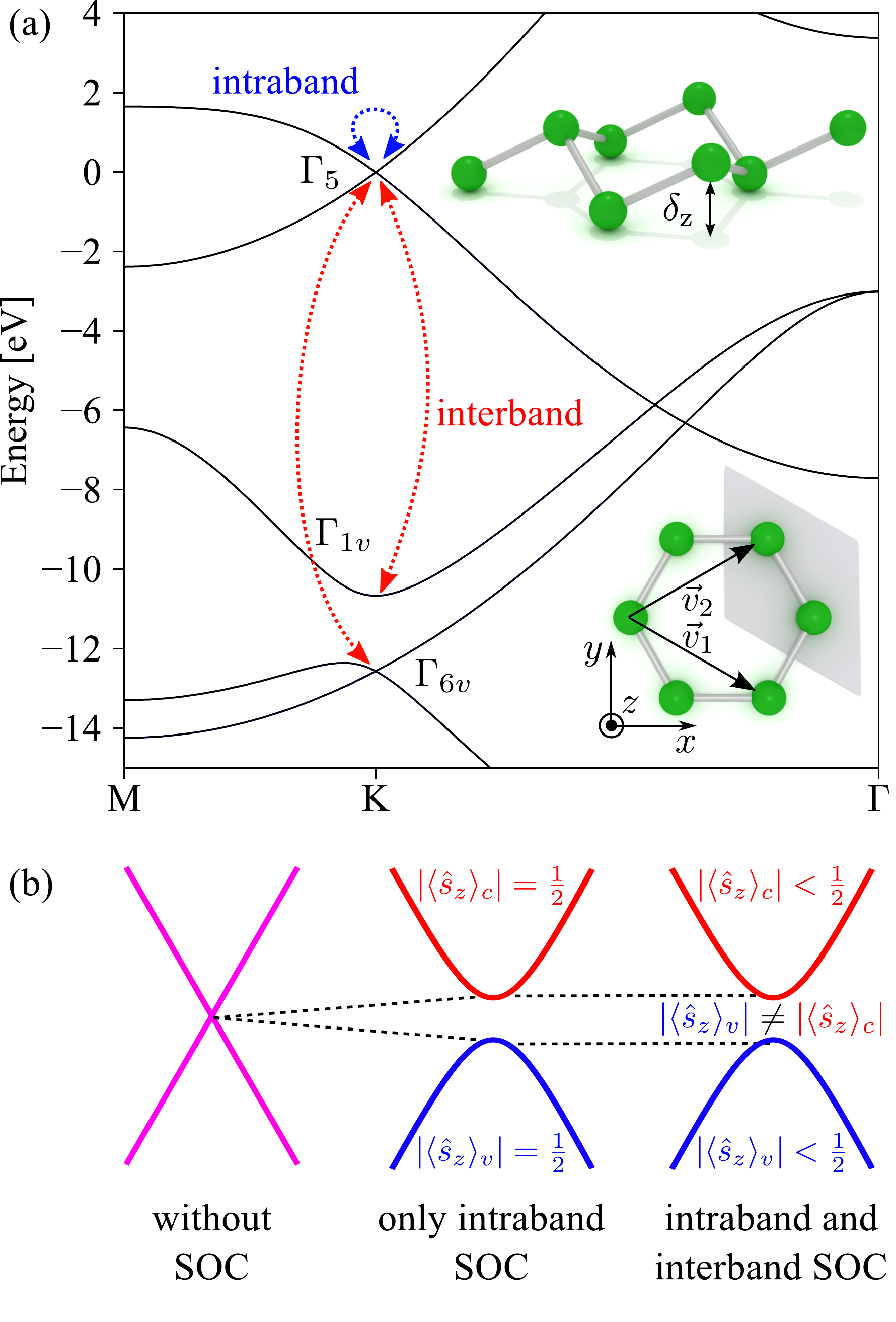}
    \caption{(Color online) 
    (a) Non-relativistic band structure of graphene with identified irreducible representations of the bands at the K-point. The labeling of the energy bands follows the irreducible representations of the $\text{D}_{3h}$ symmetry group of the K-point in graphene. Red and blue arrows visualize inter- and intraband couplings of the SOC Hamiltonian. The insets show a perspective and a top view of the crystalline structure of honeycomb 2D materials. Lattice vectors are labelled $\vec{v}_1$ an $\vec{v}_2$,  $\delta_\text{z}$ is the  out-of-plane lattice distortion ($\delta_\text{z}=0$ for graphene) and the unit cell is the shaded grey. (b) A sketch of effects of intra- and interband SO coupling on the band structure and spin expectation values $\langle \hat{s}_z\rangle$. }
    \label{fig:structure}
\end{figure}

Atomically thin two dimensional (2D) materials have been attracting attention of physicists for over a decade. Many materials have been successfully synthesized  \cite{Feng2012,vogt2012,Meng2013,Davila2014,Li2014,Derivaz2015,zhu2015nm,Zhang2016,Ji2016,Reis287,Shah2018,xia_rediscovering_2014,castellanos-gomez_isolation_2014,Shao2018,Zhang_2018} opening new routes towards novel nano-electronic and spintronic devices. 
Graphene, the first experimentally fabricated 2D material \cite{Novoselov666}, appears to be a perfect  material for spintronics \cite{Zutic2004:RMP,Fabian2007:APS} due to extraordinary long mean free path \cite{Novoselov666,Tombros2007} and weak spin-orbit coupling (SOC) of carbon atoms. 
The ideal graphene lattice is flat [$\delta_z=0$ in Fig. \ref{fig:structure} a)] and belongs to the $D_{6h}$ symmetry point group. The presence of the horizontal mirror plane of the lattice brings serious limitations to spin dynamics, as it forces the spins to be aligned perpendicularly to the graphene's plane. This is mapped into  diagonal in spin basis effective intrinsic SOC Hamiltonian in single band models \cite{Min_PRB2006}. The corresponding eigenstates are therefore pure spin \emph{up} and \emph{down} spinors, and spin scattering is prohibited. This is, however, not the full picture. Including all $p$ orbitals in the Hamiltonian leads to coupling of $\pi$ and $\sigma$ states of the opposite spins even in the presence of the mirror symmetry of the lattice\cite{fratini_2013}. But even though the states are now spin mixed, there is no effective spin scattering mechanism if graphene lattice remains flat.

The in-plane components of spin can be also present when the mirror symmetry constraint is released, as it takes place in buckled honeycomb materials such as silicene or germenene ($D_{3d}$ point group symmetry). In single band model Hamiltonians, this effect is described by the, so called, \emph{intrinsic Rashba} SOC \cite{Liu2011,Geissler2013},  or PIA SOC in the context of functionalized graphene \cite{GmitraPRL_2013,KochanPRB2017}. In contrast to well known Rashba SOC due to structure inversion asymmetry, the \emph{intrinsic Rashba} SOC does not remove the spin degeneracy of states, as a consequence of preserved space inversion symmetry.
Nevertheless, it enables the Elliott--Yafet \cite{elliott_theory_1954,Yafet_1963} (E-Y) spin relaxation mechanism, which allows spin  flips only accompanied with 
momentum scattering  by non-magnetic impurities or phonons. The latter are naturally present in  rippled structures due to flexular distortions of the lattice \cite{Mariani_2008,Mariani_erratum,Morozov_PRL_2008,Castro_PRL_2010}.

In the E-Y mechanism, the probability of spin flip follows the probability of momentum scattering, $\tau_s^{-1}\approx  b^2 \tau_p^{-1}$. \cite{elliott_theory_1954} The proportionality factor, $b^2$, is the Elliott--Yafet spin mixing (or spin admixture) parameter. It has been extensively studied for bulk materials and thin films of heavy elements \cite{Monod1979,fabian1998,steiauf2009,zimmermann2012,Long_2013,Long2013_2,zimmermann2016}, but the knowledge about $b^2$ in atomically thin 2D systems is very limited\cite{Kurpas2016, Avsar2017}.

In this paper we perform a systematic study of the SOC and spin mixing in elemental 2D materials with a honeycomb lattice structure. We 
focus 
on materials made of elements belonging to group 14 and 15 of the periodic table.  Starting from an effective multiband 
symmetry-based Hamiltonian, 
we revisit the intrinsic SOC in graphene and provide analytical solutions of the eigenstates at the K-point. We show that the expectation value of spin in the Dirac cone bands is smaller than one-half and can differ between valence and conduction band. Next, by using numerical first principles density functional theory methods we characterize intrinsic SOC and calculate spin-mixing parameter $b^2$ for graphene, silicene, germanene, stanene, blue phosphorene, arsenene, antimonene and bismuthene. We find, that the strength of the effective intrinsic SOC in the band structure $\lambda_{\text{so}}$ follows a quadratic dependence on the atomic number Z, as expected for valence electrons in isolated atoms\cite{Shanavas2014}.
The spin mixing parameter $b^2$ also follows a scaling law, $b^2 \sim Z^4$, except at spin hot spots \cite{fabian1998}. This parameter exhibits a wide range of values and giant anisotropy. 

The paper is organized as follows. In Section \ref{sec:methods}
 we briefly describe computational methods. In Section \ref{sec:theory} we discuss the effective 
 SOC Hamiltonian of graphene at 
 the K-point and show that its eigenstates are in fact mixtures of spin \emph{up} and \emph{down} states. The definition of spin-mixing parameter $b^2$ is also given here. The two forthcoming sections, Sec. \ref{sec:results} and Sec. \ref{sec:conclusions} respectively, contain numerical results for graphene, silicene, germanene, stanene, blue phosphorene, arsenene, antimonene, bismuthene with discussion and conclusions.

\section{Methods}\label{sec:methods}
The structure relaxation was prformed in {\sc Quantum ESPRESSO} package \cite{QE-2009,QE-2017}. For consistency, the PBEsol\cite{perdew_restoring_2008} exchange--interaction potential   was used for all studied  materials.  
The kinetic energy cut-offs for the wave function and charge density were individually adjusted for each element and are collected in Table \ref{relax_params}. A vacuum of 15~$\textrm{\AA}$ was introduced to avoid spurious interactions between copies of 2D films. 
Scalar-relativistic pseudopotentials were used in case of graphene and silicene, whereas for heavier elements the full relativistic pseudopotentials were applied.
The force and energy convergence thresholds for ionic minimization were set to $10^{-4}$~Ry/bohr and $10^{-5}$~Ry/bohr respectively. For the Brillouin zone integration a 12$\times$12 $k$-points mesh were generated using the Monkhorst--Pack scheme.
The optimized unit cells have been found by minimization of the total energy with respect to the lattice constant $a$. For each value of $a$ internal forces acting on atoms were relaxed using quasi--Newton scheme as implemented in  {\sc Quantum ESPRESSO}. The resulting structure parameters are collected in Table \ref{relax_params}. 

The calculations of spin properties were  performed using the all electron software package {\sc Wien2K} \cite{wien2k}. Spin--orbit coupling was included fully relativistically for core electrons, while valence electrons were treated within the second variational step method \cite{singh2006}.
Self-consistency was achieved for a 30$\times$30 $k$-points grid with 91 $k$-points in the irreducible wedge of the Brillouin zone. 

\begin{table}
\centering
\caption{\label{relax_params} Calculated lattice parameters $a=|\vec{v}_1|=|\vec{v}_2|$, $\delta_z$ [see Fig. \ref{fig:structure}], and kinetic energy cut-offs for the wave function ($E_{\text{cut}}^{\Psi}$) and charge density ($E_{\text{cut}}^{\rho}$) applied for structural optimization. The PBEsol exhange--corellation potential and 12$\times$12 $k$-point grid were assumed. }
\begin{ruledtabular}
\begin{tabular}{ lcccc}
Material  & $a$~[\textrm{\AA}] & $\delta_\text{z}$~[\textrm{\AA}]& $E_{\text{cut}}^{\Psi}$ ~[Ry]&$E_{\text{cut}}^{\rho}$~[Ry]\\
\hline
graphene & 2.459    & 0&  58 & 696 \\
silicene & 3.84& 0.45  & 58  & 580   \\
germanene & 3.99 & 0.66&  38 & 380 \\
stanene & 4.6 &0.84 & 48  & 432 \\
blue. phosp & 3.24 & 1.24 & 58 & 580\\
arsenene & 3.61& 1.38 & 45 & 450 \\
antimonene& 4.12  &1.634 &40 & 480\\
bismuthene & 4.29 & 1.73 & 42 & 429\\

\end{tabular}
\end{ruledtabular}
\end{table}


\section{Spin--orbit coupling and spin mixing}\label{sec:theory}

\subsection{Effective Hamiltonian of intrinsic SOC in graphene}

We start the analysis from the SO interaction in graphene. Being the lightest and of the highest symmetry among all the materials considered in this paper, graphene serves as a benchmark  for further discussion. In order to understand SOC effects at the Dirac point (K-point) in graphene, let us build a minimal symmetry-based Hamiltonian\cite{Winkler2010,winkler2003spin,voon2009k} by analyzing the direct coupling via the SOC operator between the Dirac cone and the nearby energy bands. 

The SOC term is given by
\begin{eqnarray}
\mathbf{H_{SO}} & = & \frac{\hbar}{4m_{0}^{2}c^{2}}\left(\vec{\nabla}V\times\vec{p}\right)\cdot\vec{\sigma}\nonumber \\
 & = & H_{SOx}\sigma_{x}+H_{SOy}\sigma_{y}+H_{SOz}\sigma_{z} \, ,
\end{eqnarray}
with the orbital components transforming as pseudovectors, i. e., $H_{SOx} \sim R_x$, $H_{SOy} \sim R_y$, $H_{SOz} \sim R_z$. Considering the energy bands of graphene shown in the Fig.~\ref{fig:structure}(a), we have the Dirac cone bands that belong to the irreducible representation (irrep) $\Gamma_5$, with states $\left|\Gamma_{5}^{1}\right\rangle \sim R_{x}$ and $\left|\Gamma_{5}^{2}\right\rangle \sim R_{y}$, the valence band $\Gamma_{6v}$ with states $\left|\Gamma_{6v}^{1}\right\rangle \sim x$ and $\left|\Gamma_{6v}^{2}\right\rangle \sim y$ and the valence band $\Gamma_{1v}$ with state $\left|\Gamma_{1v}\right\rangle \sim 1$. From the symmetry of the states and the operators within the $D_{3h}$ symmetry group of the K-point in graphene, we can find the nonzero matrix elements due to SOC, given by:

\begin{widetext}

\begin{equation}
\begin{cases}
\left\langle \Gamma_{5}^{1}\left|H_{SOz}\right|\Gamma_{5}^{2}\right\rangle =i\Delta_{5}\\
\left\langle \Gamma_{6v}^{1}\left|H_{SOy}\right|\Gamma_{6v}^{2}\right\rangle =i\Delta_{6v}\\
\left\langle \Gamma_{5}^{1}\left|H_{SOx}\right|\Gamma_{1v}\right\rangle =\left\langle \Gamma_{5}^{2}\left|H_{SOy}\right|\Gamma_{1v}\right\rangle =\Delta_{51}\\
\left\langle \Gamma_{5}^{1}\left|H_{SOx}\right|\Gamma_{6v}^{1}\right\rangle =-\left\langle \Gamma_{5}^{2}\left|H_{SOx}\right|\Gamma_{6v}^{2}\right\rangle =-\left\langle \Gamma_{5}^{1}\left|H_{SOy}\right|\Gamma_{6v}^{2}\right\rangle =-\left\langle \Gamma_{5}^{2}\left|H_{SOy}\right|\Gamma_{6v}^{1}\right\rangle =\Delta_{56} \, ,
\end{cases}
\end{equation}
with $\Delta_{5},\Delta_{6}\in\mathbb{R}$ and $\Delta_{51},\Delta_{56}\in\mathbb{C}$.

Writing the SOC Hamiltonian in the basis set
\begin{equation}
\left\{ \left[\left|\Gamma_{5}^{-}\uparrow\right\rangle ,\left|\Gamma_{1v}\downarrow\right\rangle \right],\left[\left|\Gamma_{5}^{+}\downarrow\right\rangle ,\left|\Gamma_{1v}\uparrow\right\rangle \right],\left[\left|\Gamma_{5}^{+}\uparrow\right\rangle ,\left|\Gamma_{6v}^{-}\downarrow\right\rangle \right],\left[\left|\Gamma_{5}^{-}\downarrow\right\rangle ,\left|\Gamma_{6v}^{+}\uparrow\right\rangle \right],\left[\left|\Gamma_{6v}^{-}\uparrow\right\rangle ,\left|\Gamma_{6v}^{+}\downarrow\right\rangle \right]\right\} \, ,
\end{equation}
with $\left|\Gamma_{5(6v)}^{\pm}\right\rangle =\left(\left|\Gamma_{5(6v)}^{1}\right\rangle \pm i\left|\Gamma_{5(6v)}^{2}\right\rangle \right)\big/\sqrt{2}$, we obtain the following block diagonal matrix
\begin{equation}
\left[\begin{array}{cc|cc|cc|cc|cc}
\Delta_{5} & \sqrt{2}\Delta_{51} & 0 & 0 & 0 & 0 & 0 & 0 & 0 & 0\\
\sqrt{2}\Delta_{51}^{*} & E_{1v} & 0 & 0 & 0 & 0 & 0 & 0 & 0 & 0\\
\hline 0 & 0 & \Delta_{5} & \sqrt{2}\Delta_{51} & 0 & 0 & 0 & 0 & 0 & 0\\
0 & 0 & \sqrt{2}\Delta_{51}^{*} & E_{1v} & 0 & 0 & 0 & 0 & 0 & 0\\
\hline 0 & 0 & 0 & 0 & -\Delta_{5} & 2\Delta_{56} & 0 & 0 & 0 & 0\\
0 & 0 & 0 & 0 & 2\Delta_{56}^{*} & E_{6v}-\Delta_{6} & 0 & 0 & 0 & 0\\
\hline 0 & 0 & 0 & 0 & 0 & 0 & -\Delta_{5} & 2\Delta_{56} & 0 & 0\\
0 & 0 & 0 & 0 & 0 & 0 & 2\Delta_{56}^{*} & E_{6v}-\Delta_{6} & 0 & 0\\
\hline 0 & 0 & 0 & 0 & 0 & 0 & 0 & 0 & E_{6v}+\Delta_{6} & 0\\
0 & 0 & 0 & 0 & 0 & 0 & 0 & 0 & 0 & E_{6v}+\Delta_{6}
\end{array}\right] \, ,
\end{equation}
with $E_{1v}<0$, $E_{6v}<0$, $\left|E_{1v}\right|\gg\left|\Delta_{5}\right|$,
$\left|E_{6v}\right|\gg\left|\Delta_{5}\right|$ and $\left|E_{6v}\right|\gg\left|\Delta_{6v}\right|$.
\end{widetext}

Diagonalizing the Hamiltonian, the conduction (subscript $c$) and
valence (subscript $v$) band Dirac cones have eigenvalues
\begin{align}
E_{c}& = \Delta_{5}+\frac{2\left|\Delta_{51}\right|^{2}}{\left|E_{1v}\right|+\Delta_5} \nonumber \\
E_{v}& = -\Delta_{5}+\frac{4\left|\Delta_{56}\right|^{2}}{\left|E_{6v}\right|+\Delta_{6v}-\Delta_{5}} \, ,
\end{align}
with eigenvectors
\begin{align}
\vert \psi \Uparrow \rangle_\text{c} &= \alpha\vert \Gamma_5^{-}\uparrow\rangle + \beta \vert \Gamma_{1v} \downarrow\rangle \nonumber \\
\vert \psi \Downarrow \rangle_\text{c} &= \alpha\vert \Gamma_5^{+}\downarrow\rangle + \beta \vert \Gamma_{1v} \uparrow\rangle  \\ 
\vert \psi \Uparrow \rangle_\text{v} &= \lambda\vert \Gamma_5^{+}\uparrow\rangle + \eta \vert \Gamma_{6v}^- \downarrow\rangle \nonumber \\
\vert \psi \Downarrow \rangle_\text{v} &= \lambda\vert \Gamma_5^{-}\downarrow\rangle + \eta \vert \Gamma_{6v}^+ \uparrow\rangle \nonumber \, .
\label{eq:soc_eigen}
\end{align}

The admixing coefficients are given by
\begin{align}
\alpha & =\frac{\beta\gamma}{\sqrt{2}\Delta_{51}^{*}}\nonumber \\
\beta & =\left(1+\frac{\gamma^{2}}{2\left|\Delta_{51}\right|^{2}}\right)^{-\frac{1}{2}} \\
\gamma & =\left(\left|E_{1v}\right|+\Delta_{5}\right)\left[1+\frac{2\left|\Delta_{51}\right|^{2}}{\left(\left|E_{1v}\right|+\Delta_{5}\right)^{2}}\right]\nonumber
\end{align}
and
\begin{align}
\lambda & =\frac{\eta\nu}{2\Delta_{56}^{*}}\nonumber \\
\eta & =\left(1+\frac{\nu^{2}}{4\left|\Delta_{56}\right|^{2}}\right)^{-\frac{1}{2}} \\
\nu & =\left(\left|E_{6v}\right|+\Delta_{6v}-\Delta_{5}\right)\left[1+\frac{4\left|\Delta_{56}\right|^{2}}{\left(\left|E_{6v}\right|+\Delta_{6v}-\Delta_{5}\right)^{2}}\right]\nonumber
\end{align}

From the group theory analysis we performed, it is possible to identify two different SOC contributions [depicted by arrows in Fig.~\ref{fig:structure}(a)], the intraband SOC (an interaction of the Dirac cone with itself, couples states with the same spin) and the interband SOC (the direct coupling of the Dirac cone to the valence bands $\Gamma_{1v}$ and $\Gamma_{6v}$, couples states with  opposite spins). The effect of these two SOC contributions to the Dirac cone is sketched in Fig.~\ref{fig:structure}(b). If only the intraband SOC is taken into account we notice the opening of the gap and the spin projection of the conduction (identified by the label $c$) and valence (identified by the label \textit{v}) bands of the Dirac cone is $|\left\langle \hat{s}_z \right\rangle_c| = |\left\langle \hat{s}_z \right\rangle_v| = \frac{1}{2}$. Notice that the different Dirac cone branches remain two-fold degenerate in spin and therefore it is enough to discuss the modulus of the spin projection. Adding the interband SOC contribution, the energy gap remains open but now the picture for the spin projection changes.
Due to the mixing of the energy bands via the SOC, the eigenstates of the SOC Hamiltonian at the K-point become mixtures of spin \emph{up} and \emph{down} states, given in Eq.~(\ref{eq:soc_eigen}). It immediately follows that the spin projection is reduced, $|\left\langle \hat{s}_z \right\rangle| < \frac{1}{2}$, and the two branches acquire a slightly different value of spin projection $|\left\langle \hat{s}_z \right\rangle_c| \neq |\left\langle \hat{s}_z \right\rangle_v |$. We point out, that although $|\left\langle \hat{s}_z \right\rangle| < \frac{1}{2}$ the electron's spin does not have any component along the $x$ and $y$ directions due to orthogonality of the orbital parts of the states in Eq.~(\ref{eq:soc_eigen}). The mirror symmetry of graphene is thus satisfied by the SOC Hamiltonian. The amplitudes $\beta$ and $\eta$ appear due to weak SOC, thus $|\alpha|  \gg |\beta|$ , $|\lambda|\gg |\eta|$, and one can identify  $|\beta|^2$ and $|\eta|^2$ in Eq.~(\ref{eq:soc_eigen}) as Elliott-Yafet spin-mixing parameters discussed below. It is important to
note that to obtain the correct value of the spin-orbit gap in graphene, coupling to $d$-orbital bands is needed \cite{GmitraPRB2009}. However, since the relevant $d$ states have spins perpendicular to the plane, this coupling does not contribute, to first order, to the spin mixing. From the symmetry point of view, these $d$ orbitals are already embedded in the $\Gamma_5$ states because the symmetry of the energy bands are determined from {\it ab initio}, and therefore the mixing of different orbitals are already included in the wave functions.

\subsection{Spin-mixing parameter}
Let us consider two Bloch spinors $\Psi^{\sigma}_{n,{\mathbf k}}({\mathbf r}) = u_{n,{\mathbf k}}({\mathbf r})\vert \sigma \rangle \exp({i {\mathbf k}\cdot{\mathbf r}})$, were $n$ is the band index, $u_{n,{\mathbf k}}({\mathbf r})$ is the lattice periodic function, and $\sigma=\uparrow,\downarrow$ is the electron spin.
Due to time reversal and space inversion symmetry these states are degenerate at any $\mathbf{k}$-point in the Brillouin zone (BZ), i.e., $E_n(\mathbf{k},\uparrow)=E_n(\mathbf{k},\downarrow)$.
Upon the inclusion of SOC, each of $\psi_{n,\mathbf{k}}^\sigma$ acquires an admixture of the opposite spin component forming a new pair of degenerate Bloch states
\begin{eqnarray}
\Psi^{\Uparrow}_{n,{\mathbf k}}({\mathbf r}) & = & \left[ a_{n,{\mathbf k}}({\mathbf r})\vert \uparrow \rangle + b_{n,{\mathbf k}}({\mathbf r})\vert \downarrow \rangle \right] e^{i {\mathbf k}\cdot {\mathbf r}}, \\
\Psi^{\Downarrow}_{n,{\mathbf k}}({\mathbf r}) &=& \left[ a^*_{n,-{\mathbf k}}({\mathbf r})\vert \downarrow \rangle - b^*_{n,-{\mathbf k}}({\mathbf r})\vert \uparrow \rangle \right] e^{i {\mathbf k}\cdot{\mathbf r}},
\label{eq:spinors}
\end{eqnarray} 
where $a_{n,{\mathbf k}}({\mathbf r})$ and $b_{n,{\mathbf k}}({\mathbf r})$ are again lattice periodic functions \cite{elliott_theory_1954}. 
Usually $a_{n,{\mathbf k}}({\mathbf r})$ and $b_{n,{\mathbf k}}({\mathbf r})$  are chosen in such a way, that $b_{n,{\mathbf k}}({\mathbf r})$ stands for the coefficient of the small spin component being admixed to the large spin component which has amplitude $a_{n,{\mathbf k}}({\mathbf r})$, i.e., $|b_{n,{\mathbf k}}({\mathbf r})|^2\ll |a_{n,{\mathbf k}}({\mathbf r})|^2$.
Then $\Psi^{\Uparrow}_{n,{\mathbf k}}({\mathbf r})$ is the wave function of Bloch electrons with the majority spin  \emph{up}  and  $\Psi^{\Downarrow}_{n,{\mathbf k}}({\mathbf r})$ the wave function of electrons with the majority spin \emph{down}.
Elliott pointed out  \cite{elliott_theory_1954} that the probability of a spin-flip upon momentum scattering is proportional to the  \emph{spin mixing parameter} $b^2_{n,{\mathbf k}} = \int|b_{n,{\mathbf k}}({\mathbf r})|^2 d^3{\mathbf r}$.
The analogy of $|\beta|^2$ and $|\eta|^2$ to $b^2_{n,{\mathbf k}}$ is now transparent. 

From an experimental point of view, the quantity of interest is the ensemble average of $b^2_{n,{\mathbf k}}$ for a given Fermi level rather than its value at a single ${\mathbf k}$-point. Therefore, it is useful to redefine the Elliott--Yafet spin mixing  parameter as the Fermi contour average of $b^2_{n,{\mathbf k}}$
\begin{equation}
    b^2_{\hat{\mathbf{s}}} = \frac{1}{\rho(E_F)S_{BZ}}\int_{FC}\frac{b^2_{\mathbf{k}}(\hat{\mathbf{s}})}{\hbar |\mathbf{v}_F(\mathbf{k})|}dk,
\end{equation}
where  $\hat{\mathbf{s}}$ is the unit vector defining the spin quantization axis (SQA), $S_{BZ}$ is the area of the Fermi surface, $\rho(E_F)$ is the density of states per spin at the Fermi level, $\mathbf{v}_F(\mathbf{k})$ is the Fermi velocity and the integration takes over an iso-energy contour.  In electrical spin injection experiments SQA corresponds to the polarization of initial magnetization of populated electrons. Such a definition allows us to
explore the anisotropy of $b^2$ in the band structure, similarly to what was done for selected 3D materials \cite{zimmermann2012,zimmermann2016}.


\section{Results and discussion}\label{sec:results}
The initial structure parameters of silicene, germanene, and stanene have been taken from Ref. [\onlinecite{Matusalem2015}]. For arsenene we used parameters from Ref. [\onlinecite{Kamal2015}], and for graphene we used the initial lattice constant 2.46~\textrm{\AA}. Optimized lattice parameters and buckling heights are very close to the original values and are listed in Table \ref{relax_params}.

\begin{figure*}[ht]
    \centering
     \includegraphics[width=0.9\textwidth]{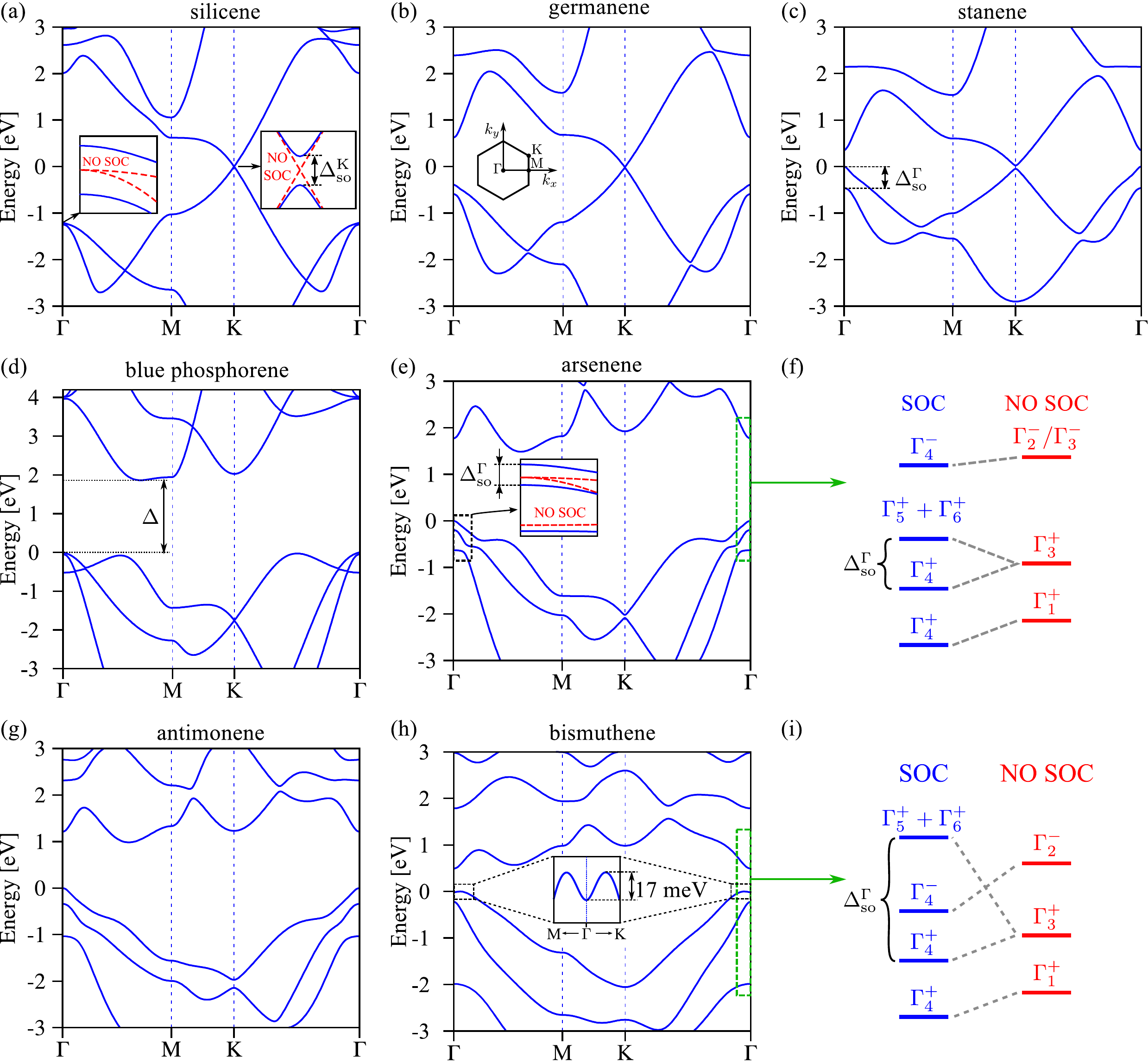}
    \caption{(Color online) Relativistic band structures from first--principles plotted along high symmetry points of the first Brillouin zone, shown as the inset in (b).
    The insets in (a) and (e) visualize the splitting of degenerate orbital states at the $\Gamma$ ($\Delta_{\text{so}}^\Gamma$) and K ($\Delta_{\text{so}}^\text{K}$) points upon the inclusion of spin--orbit coupling. The corresponding values of $\Delta_{\text{so}}^\Gamma$ and $\Delta_{\text{so}}^\text{K}$ are collected in Table \ref{tab_gaps}. (f) Ordering of four bands close to the band gap at the $\Gamma$-point ($D_{3d}$ symmetry group) with (blue) and without (red) SOC for blue phosphorene ($\Gamma_3^-$ irrep for the conduction band without SOC), arsenene and antimonene ($\Gamma_2^-$ irrep for the conduction band without SOC). (h) Same as is (f) but for bismuthene. The strong SOC induces crossing of the two top-most bands and leads to band gap inversion.}
    \label{fig:bandstructures}
\end{figure*}

\paragraph{Spin-orbit splitting.}
At first we focus focus on materials from group 14.  The band structure of graphene is shown in Fig.  \ref{fig:structure} and was discussed above. In the top row of Fig. \ref{fig:bandstructures} we show the calculated relativistic band structures of silicene, germanene, and stanene.
All these materials are semimetals.
The semimetalic character is manifested by the presence of a Dirac cone
centered in the Brillouin zone at the K-point. Without SOC the valence and conduction branches of the cone
touch at the Fermi energy, forming a zero-width band gap [dashed red line in the insets of Fig. \ref{fig:bandstructures} (a)].
The inclusion of SOC pulls them apart and introduces a spin-orbital gap $\Delta_{\text{so}}^{\text{K}}$, while the spin degeneracy is preserved by virtue of time reversal and space inversion symmetry. For graphene, silicene, and germanene, the spin-orbital gap $\Delta_{\text{so}}^{\text{K}}$ is synonymous with the fundamental band gap $\Delta$, 
defined as the energy distance between the valence and conduction band edges [Fig. \ref{fig:bandstructures} (d)].
At the $\Gamma$-point, SOC splits off the two top--most valence bands, by the energy $\Delta_{\text{so}}^\Gamma$ [Fig. \ref{fig:bandstructures} (a),(c)], and other bands lying far away from the Fermi level and being irrelevant to the discussion of low energy physics we focus on here. 
For graphene, silicene, and germanene the split-off bands at the $\Gamma$-point lie below the valence band maximum (VBM) at the K-point.   
For stanene [Fig. \ref{fig:bandstructures} (c)], due to strong SOC, $\Delta_{\text{so}}^\Gamma = 200$~meV, the energy of the top-most valence band at the $\Gamma$-point overtakes the energy at the K-point by 47~meV and the band gap becomes indirect.

Contrary to group 14 semimetals, materials made of group 15 elements  are semiconductors with sizable indirect band gaps [Fig. \ref{fig:bandstructures} (d),(e),(g),(h)]. The calculated values of the band gaps $\Delta$ are: 1.86~eV, 1.49~eV,  0.98~eV and 0.49~eV for blue phosphorene, arsenene, antimonene and bismuthene, respectively (black phosphorene was investigated in Ref.~\onlinecite{Kurpas2016}).
These values are consistent with other calculations \cite{Kamal2015,zhu2015,Kecik2016}, and with experimental reports \cite{Shah2018}.
Except for bismuthene, the VBM is located at the $\Gamma$-point, while the conduction band minimum (CBM) lies close to the middle of $\Gamma$--M path in the Brillouin zone. The spin-orbital splittings of two top-most valence bands at the $\Gamma$-point $\Delta_{\text{so}}^\Gamma$ are 48~meV for blue phosphorene, 195~meV for arsenene and 340~meV for antimonene. 
Bismuthene [Fig. \ref{fig:bandstructures}(h)] displays qualitatively different picture due to the inverted band gap. 
The gap inversion can be easily  identified by comparing the band ordering for bismuthene and the remaining materials of group 15. In Fig. \ref{fig:bandstructures}(i), we show irreducible representations of the four relevant bands of bismuthene. 
Without SOC the top-most valence band $\Gamma^+_3$ is two-fold degenerate. 
\begin{table}
\centering
\caption{\label{tab_gaps} Orbital ($\Delta$) and spin--orbital gaps ($\Delta_{\text{so}}^{\text{K}}$), ($\Delta_{\text{so}}^{\Gamma}$) calculated in {\sc Wien2K} for PBEsol exchange--correlation potential. The character of the orbital gap, direct or indirect, is labelled by capital the letter D or I respectively. Inverted band gaps are indicated by Inv. }
\begin{ruledtabular}
\begin{tabular}{lccc}
Material  &$\Delta$~[meV]&$\Delta_{\text{so}}^{\text{K}}$~[meV]&$ \Delta_{\text{so}}^{\Gamma}$~[meV]\\
\hline  
graphene  & 0.024~(D) & 0.024 & 9 \\
silicene  & 1.48~(D)  & 1.48  & 34.6  \\
germanene & 23~(D)    & 23    & 201 \\
stanene   & 25~(I)    & 72    & 461 \\
blue phosphorus & 1864~(I)& 10 & 48 \\
arsenene  & 1492~(I)& 71 & 195 \\
antimonene & 982 (I)& 174& 340 \\
bismuthene & 491~(I, Inv)&702 & 712 \\
\end{tabular}
\end{ruledtabular}
\end{table}
Upon inclusion of SOC it splits off into two bands $\Gamma^{+}_4$ and $\Gamma^+_5 + \Gamma^+_6$ separated by $\Delta_{\text{so}}^\Gamma \approx 700$ ~meV [see Fig. \ref{fig:bandstructures} (i)]. The latter band interchanges with the first conduction band $\Gamma_4^-$ and the gap becomes inverted, with respect to band ordering of lighter materials of group 15 [see Fig. \ref{fig:bandstructures}~(f)].
The  edge of the valence band of bismuthene lies slightly away from the $\Gamma$-point [see inset in Fig. \ref{fig:bandstructures}(h)], with energy only 17~meV higher than the energy of the band at the $\Gamma$-point, and the character of the band gap is almost direct.
For all group 15 materials the characteristic Dirac cone lies approximately 2~eV below the valence band maximum and gradually loses its  linear dispersion character with an increasing atomic number Z. \\
\begin{figure}
    \centering
    \includegraphics[width=\columnwidth]{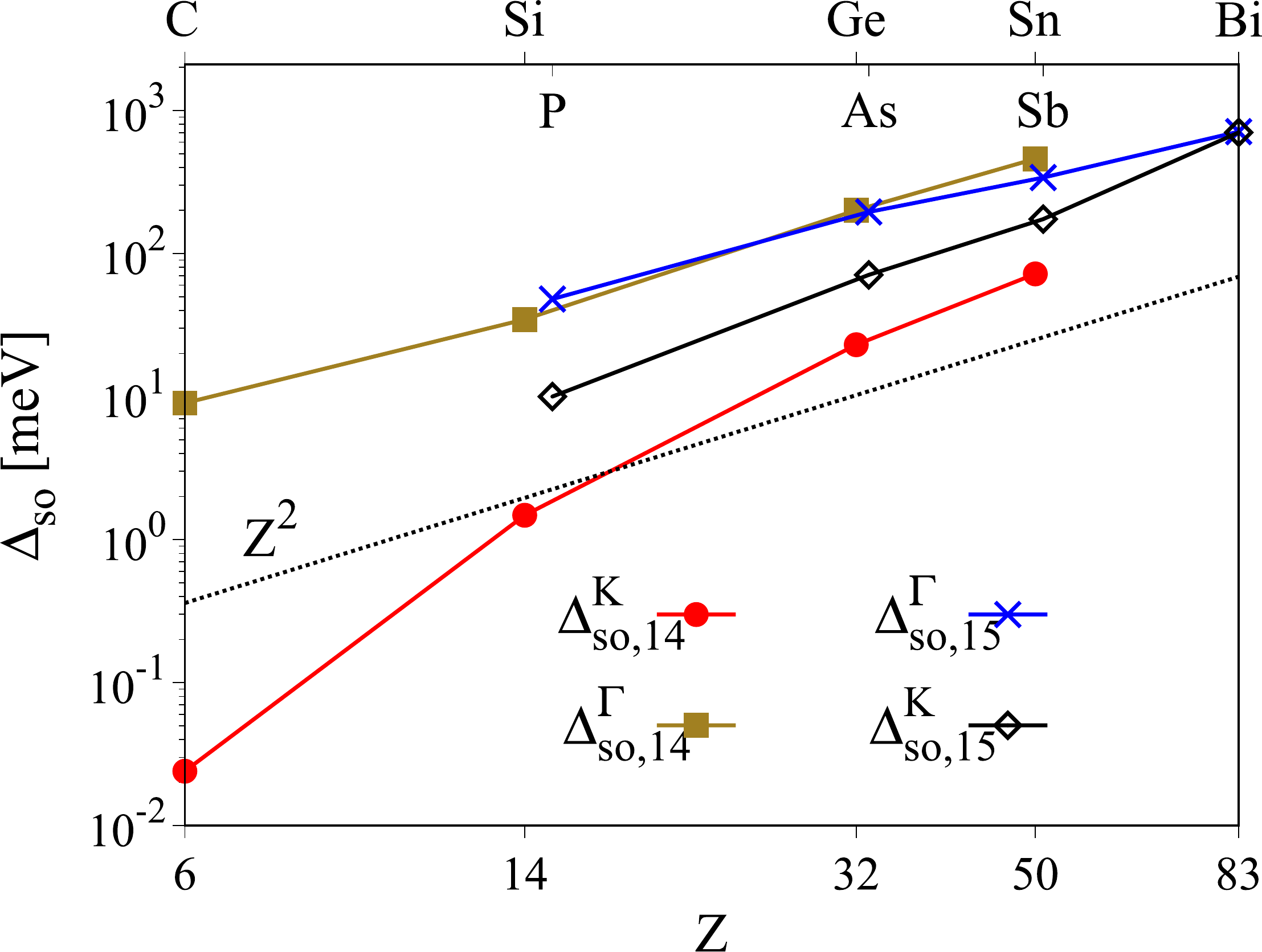}
    \caption{\label{fig:delta_b2k_K}(Color online)  Spin--orbital gap at the K ($\Delta_{\text{so}}^\text{K}$) and $\Gamma$ ($\Delta_{\text{so}}^{\Gamma}$) points versus the atomic number Z for materials of group 14 ($\Delta_{\text{so,14}}^\text{K}$, $\Delta_{\text{so,14}}^{\Gamma}$) and of group 15 ($\Delta_{\text{so,15}}^\text{K}$, $\Delta_{\text{so,15}}^{\Gamma}$). The names of elements are shown on the top \textit{x}-axis.
    A quadratic function of Z is plotted for the reference (dotted black line). 
    }
\end{figure}

Since all the studied materials have the same crystalline structure one can expect that, within the same group of periodic table, the spin-orbital gap $\Delta_{\text{so}}$ will mainly depend on the electronic configuration of the element. In the first order perturbation theory $\Delta_{\text{so}} \sim \lambda_{\text{so}}$, where $\lambda_{\text{so}}$ is the strength of SOC in the band structure. In isolated atoms, if only valence electrons are taken into account,  $\lambda_{\text{so}}\sim Z^2$. \cite{Shanavas2014} In crystalline solids, bands close to the Fermi level are made of states of valence electrons. Therefore, one can roughly expect that $\Delta_{\text{so}} $ will also follow a quadratic dependence on Z.
On the other hand, the effective SO interaction in a band is, generally, momentum dependent, and includes contributions from other bands coupled by the SO interaction. 
A systematic study of SO interaction would be necessary to visualize the global behavior, but this goes beyond the scope of this paper. Instead, we will focus only on the high symmetry points K and $\Gamma$. 
In Fig. \ref{fig:delta_b2k_K} we plotted $\Delta_{\text{so}}^{\text{K}}$ and $\Delta_{\text{so}}^{\Gamma}$ as a function of atomic number Z.  The values are collected  in Table \ref{tab_gaps}.
It is seen that indeed $\Delta_{\text{so}}$ follows the $Z^2$ dependence very well. A deviation from the common quadratic dependence is seen for graphene and silicene at the K-point. This may be caused by the fact that the core potential is not effectively screened due to low number of core electrons.
The deviation for graphene is explained by the absence of buckling and therefore of scalar coupling between $p_z$ and in-plane orbitals; that is, the prefactor of the
scaling is drastically reduced.
\begin{figure}[ht]
    \centering
    \includegraphics[width=\columnwidth]{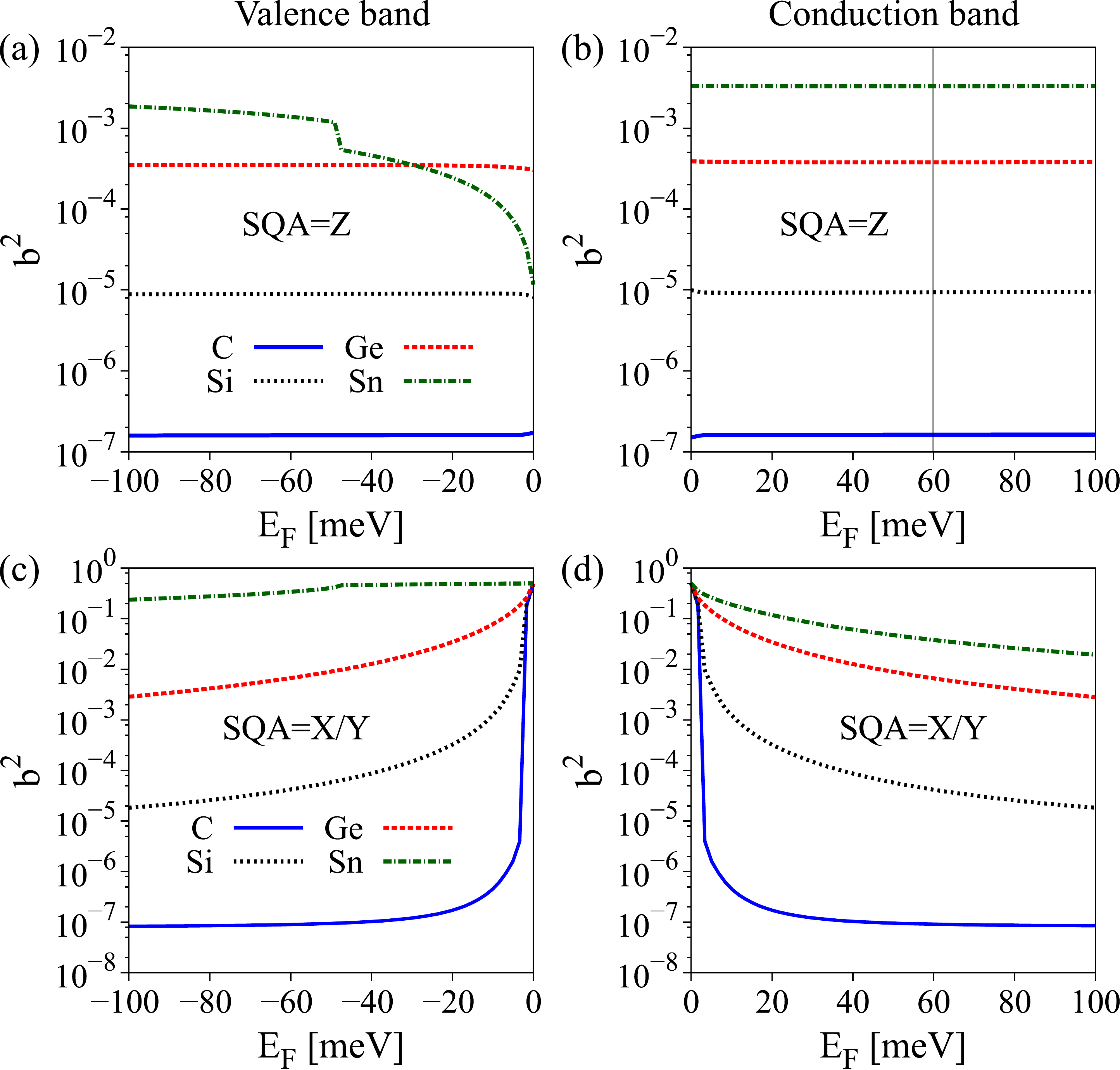}
    \caption{(Color online) Calculated average spin--mixing parameter $b^2$ versus Fermi energy relative to the valence (conduction) band maximum (minimum) for materials of group 14. Materials are labelled by the element name: C - graphene, Si - silicene, Ge - germanene, Sn - stanene. (a) Valence band and SQA=Z (b) Conduction band and SQA=Z. The solid grey vertical line marks the values of $b^2$ plotted in Fig \ref{fig:aver_b2_scaling}. (c) Same as (a) but for SQA=X/Y. (d) Same as (b) but for  SQA=X/Y. \label{fig:b2k_14}}
 
\end{figure}

\paragraph{Spin-mixing.}

Let us now discuss the spin--mixing parameter $b^2$. In Fig. \ref{fig:b2k_14} we show the calculated Fermi contour averaged spin--mixing parameter for group 14 materials. 
For out-of-plane spin polarization (SQA=Z), $b^2$ is almost independent of the position of the Fermi energy $\text{E}_\text{F}$, both for the valence and conduction band. 
We relate it to the fact, that around the K-point the two bands forming the Dirac cone are well separated from the others and the effective SOC in the valence and conduction bands near the K-point is almost momentum independent (within the range of doping considered here). The intraband SOC, involving the valence and conduction branch of the cone, does not contribute to $b^2$, as was shown by our effective model. The intrinsic Rashba (PIA) SOC vanishes at the K-point  and grows linearly with momentum \cite{Liu2011,Geissler2013,GmitraPRL_2013,KochanPRB2017}. Therefore its contribution around the K-point is small. For stanene [Fig. \ref{fig:b2k_14}~(a)] the valence band edge is at the $\Gamma$-point. 
Initially very small value of $b^2$ rapidly increases with doping, due to interaction with the lower valence band. At $\text{E}_\text{F}\approx -47$~meV the band around the K-point starts contributing to the Fermi contour and a discontinuous increase of $b^2$ is observed. We stress, that the finite value of $b^2$ for graphene, of the order of $10^{-7}$, does not imply
spin-flip scattering by scalar impurities. Such scattering is prohibited by the mirror symmetry of the lattice, i.e. $\langle \Gamma_5^{\pm}|V_{\rm imp}|\Gamma_{1v}\rangle=\langle \Gamma_5^{\pm}|V_{\rm imp}|\Gamma_{6v}^{\mp}\rangle=0$, if the impurity potential $V_{\rm imp}$ is even upon  mirror reflection.

For spins polarized in-plane (SQA=X/Y,  bottom row in Fig.~\ref{fig:b2k_14}) $b^2$ is almost one-half for $\text{E}_\text{F}=0$ due to the spin hot spot \cite{fabian1998} at the K-point (only $z$ component of spin is allowed). 
With increasing doping it starts to decrease towards the values similar to  SQA=Z. Again, $b^2$ in the valence band of stanene is an exception. For the whole doping range it does not go below $b^2= 0.2$, and spins remain almost fully mixed. Even for high doping,  $\text{E}_\text{F}=-100$~meV, $b^2$ for in-plane spin orientation is two orders of magnitude greater than for out-of-plane spins. 

\begin{figure}
    \centering
    \includegraphics[width=\columnwidth]{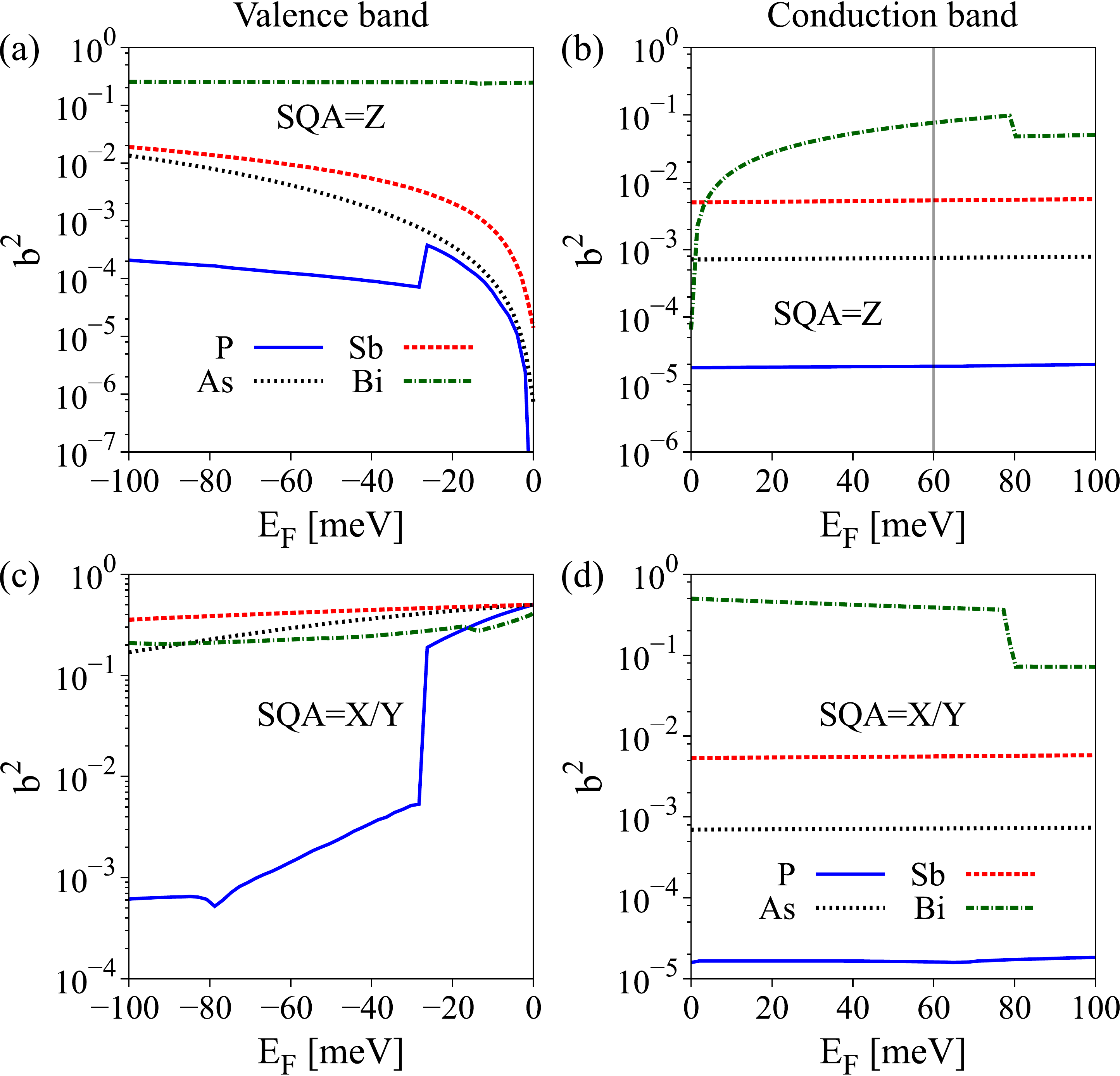}
    \caption{(Color online) Calculated average spin--mixing parameter $b^2$ versus Fermi energy relative to the valence (conduction) band maximum (minimum) for materials of group 15. Materials are labelled by the element names: P - blue phosphorene, As- arsenene, Sb- antimonene and Bi - bismuthene. (a) Valence band and SQA=Z (b) Conduction band and SQA=Z. The solid grey vertical line marks the values of $b^2$ plotted in Fig \ref{fig:aver_b2_scaling}.  (c) Same as (a) but for SQA=X/Y. (d) Same as (b) but for  SQA=X/Y. \label{fig:b2k_15}}
\end{figure}

For group 15 materials [Fig. \ref{fig:b2k_15}] $b^2$ displays more diversity due to more complicated band structures around the band gap. 
Nevertheless, similar trends as for group 14 materials can be identified: 
(i) for SQA=Z, if the VBM is centered at the $\Gamma$-point, $b^2$ grows exponentially  when moving away from the Brillouin zone center. This happens in the valence band of blue phosphorene, arsenene, and antimonene as shown in Fig. \ref{fig:b2k_15} (a), or in the conduction band of bismuthene [Fig. \ref{fig:b2k_15} (b)].
(ii) When the band edge is away from the high symmetry points and the band is relatively well separated from the others (no spin hot spots due to accidental anticrossings occur), $b^2$ exhibits a small variation with doping. Such behavior is observed in the conduction band of phosphorene, arsenene, and antimonene  [Fig. \ref{fig:b2k_15} (b),(d)].
The reduced symmetry in k-space also results in weak anisotropy of $b^2$. The values of $b^2$ in the conduction and valence bands of phosphorene, arsenene and antimonene, are almost identical.
(iii) A discontinuous change of $b^2$ takes place when a next band crosses the Fermi level. The contribution of this band to the averaged $b^2$ is far from being trivial. It depends on the form  and strength of SOC in the band at a given $\mathbf{k}$-point, and on the number of states contributing to the Fermi contour. For example, for stanene $b^2$ decreases when the energy band around the K-point starts contributing to the total average [Fig. \ref{fig:b2k_14} (a)], and for  phosphorene decreases when another valence band crosses the Fermi level [Fig. \ref{fig:b2k_15} (a),(c)].

\begin{figure} 
    \centering
    \includegraphics[width=\columnwidth]{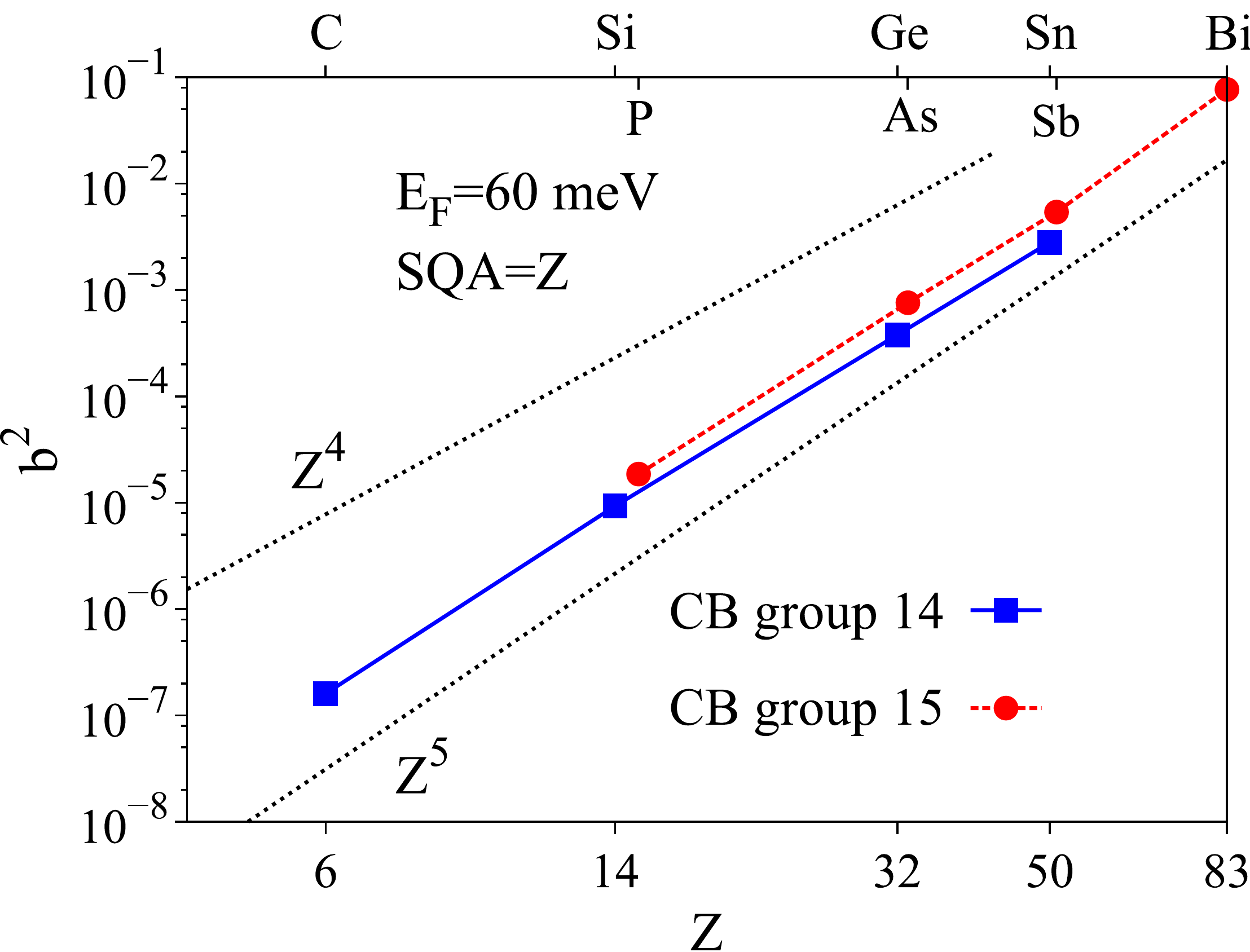}
    \caption{(Color online) Averaged spin-mixing parameter $b^2$ in the conduction band and SQA=Z versus the atomic number Z. The names of elements are shown on the top \textit{x}-axis. The values of $b^2$ were taken from Figs \ref{fig:b2k_14}~(b) and \ref{fig:b2k_15}~(b) at $\text{E}_\text{F}=60$~meV (marked by vertical lines in the corresponding figures).  \label{fig:aver_b2_scaling}}
\end{figure}

We have also checked how $b^2$ scales with the atomic number Z. 
Within first order non-degenerate perturbation theory, the admixture amplitude  in Eq.  (\ref{eq:spinors}), $b_{n,{\mathbf k}}$ is proportional to $\lambda_{\text{so}}$.
Taking $\lambda_\text{so} \sim \text{Z}^2$ one can expect  that $b^2$ should follow $\text{Z}^4$ dependence.
 In Fig. \ref{fig:aver_b2_scaling} we plot average $b^2$ in the conduction band versus the atomic number Z (corresponding to a given material) for iso-energy contour at $\text{E}_\text{F}=60$~meV and SQA=Z [see grey vertical lines in Figs \ref{fig:b2k_14}~(b) and \ref{fig:b2k_15}~(b)]. The Fermi energy was chosen such that the spin-mixing parameter is not strongly influenced by the vicinity of a spin hot spot and reflects the pure SOC in the band. As can be seen, our results agree well with the estimate given by perturbation theory, though small deviation from $Z^4$ are observed. This can be attributed to  a rather complex nature of spin-orbit coupling in many electron crystalline solids. 

\begin{figure}
    \centering
    \includegraphics[width=0.99\columnwidth]{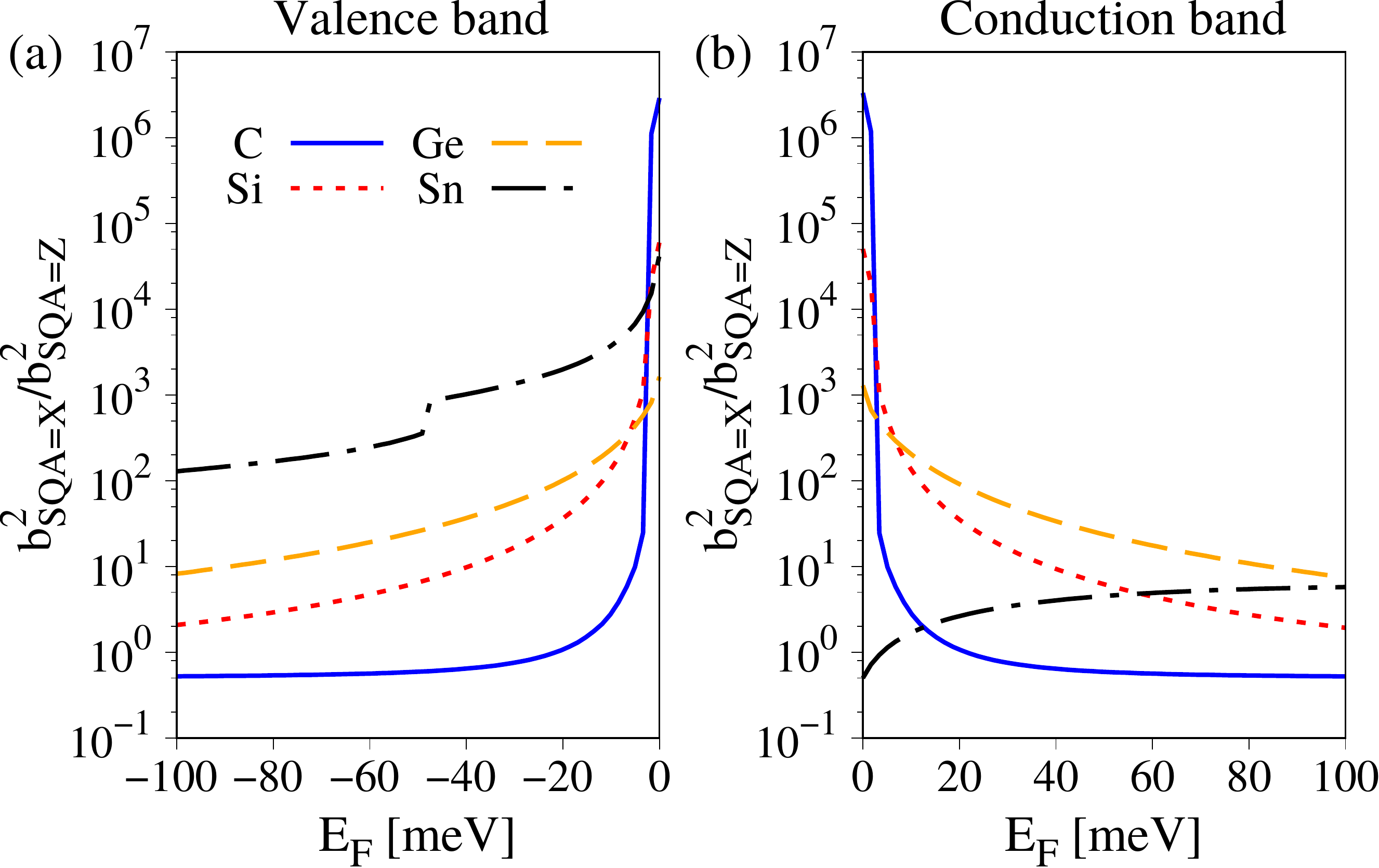}
    \caption{\label{fig:b2_anisotropy}(Color online)
    Anisotropy of spin mixing parameter $b^2_{\text{SQA=X/Y}}/b^2_{\text{SQA=Z}}$ versus Fermi  energy for materials made of elements of group 14. (a) valence band, (b) conduction band. The Fermi energy is given with respect to the valence band maximum. 
    }
\end{figure}

\begin{figure}
    \centering
    \includegraphics[width=\columnwidth]{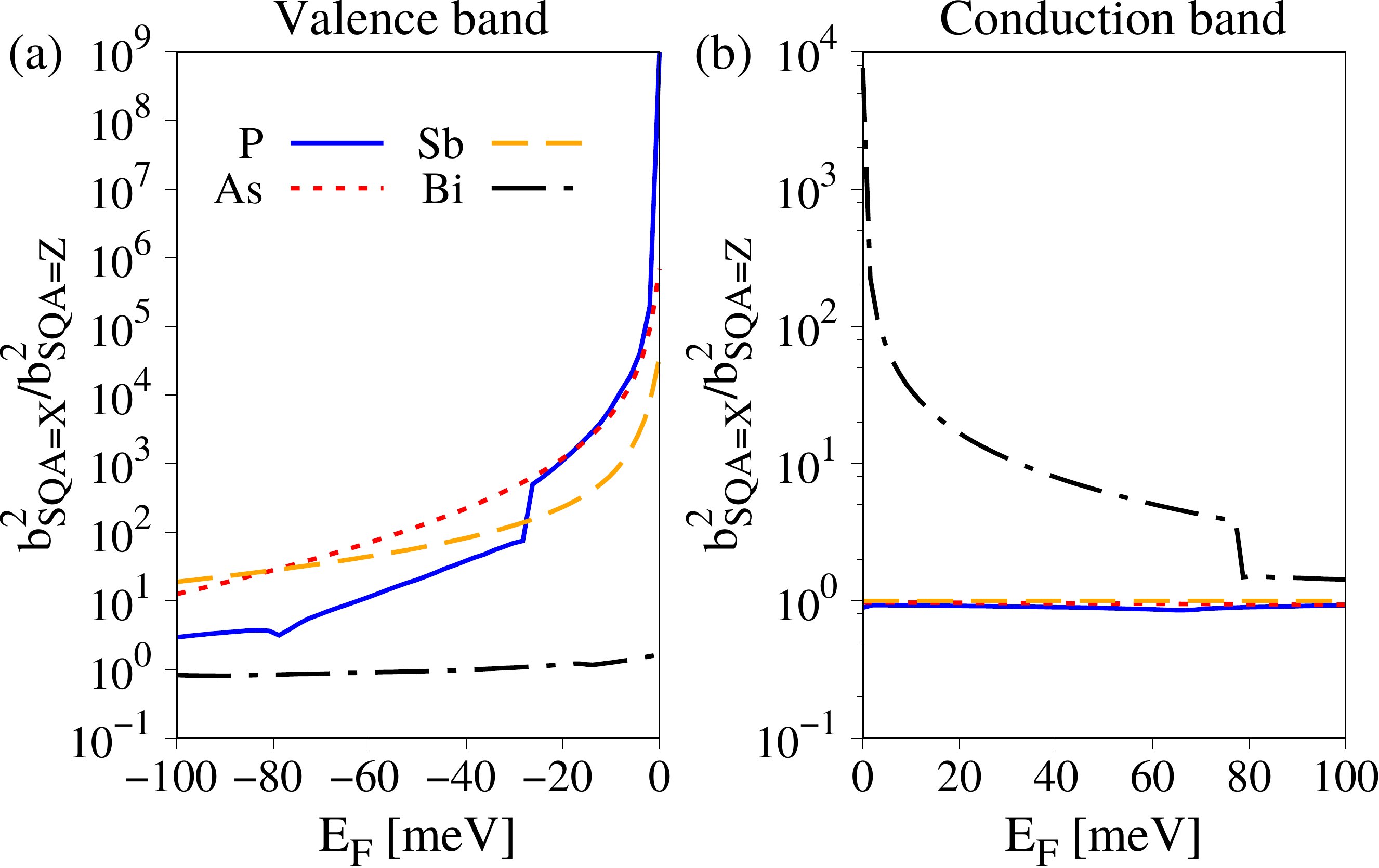}
    \caption{(Color online) 
    Anisotropy of spin mixing parameter $b^2_{\text{SQA=X/Y}}/b^2_{\text{SQA=Z}}$ versus Fermi  energy for materials made of elements of group 15. (a) valence band, (b) conduction band. The Fermi energy is given with respect to the valence band maximum. 
    \label{fig:b2k_anis_cond}}
\end{figure}

Finally, we have calculated spin mixing anisotropy, which is a measure of spin relaxation anisotropy. The ratio $b^2_{\text{SQA=X}}/b^2_{\text{SQA=Z}}$ (in-plane to out-of-plane spin polarization) for materials of group 14 is shown in Fig. \ref{fig:b2_anisotropy}. All materials display giant (when compared to corresponding anisotropies of 3D materials) and doping dependent anisotropy. For most  materials, the anisotropy is driven by the spin hot spot for in-plane polarized spins at the K or $\Gamma$ points. 
The highest anisotropy 
at large  $\text{E}_\text{F}$ 
is observed for stanene in the valence band, between $10^{2}$ and $10^3$, and results in strong spin mixing for in-plane spin polarization. For graphene, we find that $b^2_{\text{SQA=X}}/b^2_{\text{SQA=Z}} \approx 0.5$ for $E_F > 30$~meV.
Similar trends are observed for materials of group 15. In the valence band [Fig. \ref{fig:b2k_anis_cond}~a)] a spin hot spot at the $\Gamma$ point ($E_F=0$~meV) results in a huge anisotropy, which decreases when moving away from the high-symmetry point. This happens for blue phosphorene, arsenene, and antimonene. Bismuthene displays almost no anisotropy of $b^2$ in the valence band due to strong spin mixing for all spin polarizations. The picture is oposite in the conduction band [Fig. \ref{fig:b2k_anis_cond}~b)]. Anisotropic behavior of $b^2$ is found for blue phosphorene, arsenene, and antimonene, while $b^2$ for bismuthene shows doping dependent anistropy.
 Essentially, $b^2$ exhibits a strong anisotropy if the BZ wedge defined by the doping range contains spin hot spots or spin hot regions, while the anisotropy is not well pronounced  otherwise. 

The above calculated spin admixture can be used to obtain realistic estimates of spin relaxation times. Indeed, all the studied elemental 2D materials have space inversion symmetry and are thus expected to exhibit spin relaxation according to the Elliott-Yafet mechanism \cite{elliott_theory_1954,Yafet_1963}. The only other input needed is the momentum relaxation time $\tau_p$ which can be obtained from electrical transport, for example. 
The link to spin relaxation is provided by the
Elliott relation $\tau_s^{-1}\approx  b^2 \tau_p^{-1}$, which should be valid in regions where $b^2 \alt 0.2$, where perturbation theory
holds. Two connected basic assumptions are needed: the spin-orbit coupled bands should be spectrally separated by more than is the spin-orbit coupling
matrix element between them, and the spin expectation value is close to 1/2.\cite{elliott_theory_1954} 

At spin hot spots, which occur at $K$ and $\Gamma$ points for our studied materials, these two assumptions are in general violated and the mechanism needs to be modified, see for example Ref. [\onlinecite{Boross2013,szolnoki2017}]. At these points, for the spin direction at which $b^2$ is of order 1, spin relaxation and and momentum relaxation times become comparable \cite{FabianPRL83}.

\section{Conclusions}\label{sec:conclusions}
We have performed a systematic study of spin-orbit coupling in elemental two-dimensional materials of group 14 and 15 of the periodic table. Starting from symmetry arguments we have formulated an effective 
multiband symmetry-based SOC 
Hamiltonian for graphene at the K-point. We have shown that even if the mirror symmetry of the lattice protects the spin in graphene from acquiring the $x$ and $y$ components, spin mixing due to the intrinsic SOC is still possible but does not lead to spin relaxation.
Using first principles numerical methods we analyzed intrinsic SOC and calculated the Elliott-Yafet spin-mixing parameter $b^2$ for graphene and other honeycomb lattice materials. We have shown that spin-orbit coupling in the band structure scales as a square function of the atomic number Z. Away from spin hot spots the spin-mixing parameter also follows the exponential scaling power law, $b^2\sim \text{Z}^4$.
We identified three main factors having the strongest influence on the overall average value of $b^2$, namely,  the strength of the intrinsic SOC, the shape of the Fermi contour, and the presence of spin hot spots inside or close to the contour. For almost all materials $b^2$ shows substantial and doping dependent anisotropy. Our results for $b^2$ can be translated into spin relaxation times, once the momentum relaxation time is known. Therefore they provide valuable information on the potential application of those materials in spintronics. 


\section*{Acknowledgments}\label{sec:sck}
This work was supported by the National Science Centre under the contract DEC-2018/29/B/ST3/01892, Alexander von Humboldt Foundation, Capes (grant No. 99999.000420/2016-06), SFB 1277 (A09 and B05) in part by PAAD Infrastructure co-financed by Operational Programme Innovative Economy, Objective 2.3, and by the Scientific Grant Agency of Ministry of Education of Slovak Republic (Grant No. 1/0531/19) and VVGS-2018-887. The authors gratefully acknowledge the Gauss Center for Supercomputing e.V. for providing computational resources on the GCS Supercomputer SuperMUC at the Leibniz Supercomputing Center.

\bibliography{references.bib}
\end{document}